\documentstyle[preprint,aps,prl,epsf]{revtex}
\begin{document}
\def\bea{\begin{eqnarray}}
\def\eea{\end{eqnarray}}
\def\a{\alpha}
\def\d{\delta}
\def\p{\partial} 
\def\nn{\nonumber}
\def\r{\rho}
\def\rv{\bar{r}}
\def\la{\langle}
\def\ra{\rangle}
\def\e{\epsilon}
\def\o{\omega}
\def\n{\eta}
\def\g{\gamma}
\def\break#1{\pagebreak \vspace*{#1}}
\def\f{\frac}
\draft
\title{Biopolymer Elasticity}
\author{Supurna Sinha}
\address{Raman Research Institute,
Bangalore 560080,India\\}
\maketitle
\widetext
\begin{abstract}
In recent years molecular elasticity has 
emerged as an active area of research:
there are experiments that probe mechanical properties of
single biomolecules such as DNA and
Actin, with a view to understanding the role of elasticity of 
these polymers in biological 
processes such as transcription and protein-induced DNA bending. 
Single molecule elasticity has thus
emerged as an area where there is 
a rich cross-fertilization of ideas between biologists,
chemists and theoretical physicists. In this article we present 
a perspective on this field of research.

\end{abstract}
\newpage
\narrowtext

Polymers of different flexibilities are vital to life.
Some polymers are as stiff as needles, others are 
as flexible as thread and still others are like twine, neither too stiff 
nor
too floppy. 
The cytoskeletal structure of the cell, which gives rigidity to the 
cell, consists of semiflexible polymers - like Actin, Microtubules and 
intermediate filaments. The polymer which carries the genetic code,
the DNA, is 
a semiflexible polymer (See Box $1$). A typical human DNA in a cell is about a meter
long and it is rather amazing that such a long molecule is packed into
a cell nucleus of the size of a few microns. This efficient packaging
is a mystery and raises questions about the elasticity of DNA. 
DNA also assumes various 
different forms depending on the function it needs to perform
- for instance, it is unwound 
during replication so that the genetic code can be accessed and copied.
At other times the DNA assumes a supercoiled configuration which protects
and preserves the genetic code intact till it is needed. 
Such conformationally distinct configurations are also
dictated by its elasticity.    

{\bf Single Molecule Experiments:}
\vskip .2cm

Experimental studies of biopolymer molecules such as
DNA have traditionally been limited to samples
containing large numbers of molecules. 
This made it hard to probe the
elastic properties of individual biopolymers which are of
vital importance to biological
processes such as protein-induced DNA bending.
It is only quite recently, due to advances in technology
that single molecule studies became feasible.
It is now possible to design experiments in which single molecules are
pulled, stretched and twisted 
to measure elastic
properties.
In a typical single molecule experiment, one considers a polymer 
molecule suspended between a fixed surface and a force sensor of some 
kind (See Fig. $1$). 
The force sensor is a bead in a laser trap or a flexible cantilever of 
an atomic force microscope and the molecule under consideration can range
from a DNA in a simple random coil configuration to a globular protein 
in a unique three-dimensionally folded form. 
For instance, one can study the
``equation of state'' of a 
semiflexible polymer by measuring its extension as a 
function of applied
force. One can also tag the ends with fluorescent dye
and
determine the distribution
of end-to-end distances. We learn a lot from such experimental studies 
about biologically relevant mechanical 
properties of these polymers. 

Such single molecule micromanipulation studies have opened up a
vast area of cross-fertilization of ideas between biologists, chemists
and theoretical physicists. 
Results of single molecule experiments have posed challenges
to theoretical physicists who have combined ideas from statistical mechanics,
quantum mechanics, differential geometry and topology to get a proper
understanding of the field (See Box $2$). 

{\bf Force-Extension Curves: Experiment And Theory}
\vskip .2cm
Carlos Bustamante and his colleagues directly measured the elasticity
of a single DNA molecule by attaching one end of the molecule to a 
glass slide and the other end to a micron size magnetic bead. A stretching
force was applied to the DNA molecule by applying a magnetic field 
gradient and the 
corresponding extension was measured by observing the position of the bead
under an optical microscope.
The mechanical response of such a system was thus probed
in the form of a force-extension
curve.

The experimental curves were compared first against the simplest 
theory available for a polymer- the freely jointed chain (FJC) in which 
the polymer is viewed as a chain of rigid rods connected by revolving 
pivots. While the experimental curves agreed well with the FJC model
at low forces, the 
comparison showed considerable discrepancies between theory
and experimental curves in the intermediate and large force regime. 
This pointed towards a deficiency in the 
theory which was subsequently remedied  by Marko and Siggia who
used the Worm Like Chain (WLC) model. 
In the WLC model one takes into consideration the bending energy 
(${\cal E}= \frac{1}{2}A\int_0^L ds\kappa^2$
with $A=L_p k_BT$ the  bending modulus)
proportional to the square of the curvature $\kappa$
of the space curve representing the polymer of contour length $L$ kept
at a temperature $T$.
The WLC model 
provided excellent quantitative agreement with the
experimental force-extension curves (See Fig. $2$).  
This pointed to the fact that a DNA molecule needs to be viewed as 
a semiflexible (partly flexible and partly stiff) polymer rather than a 
completely flexible one.

{\bf Twisting Single Molecules:}

In the experiments of Strick and collaborators
the ends of a single molecule of double stranded DNA are attached to
a glass plate and a magnetic bead.
Magnetic fields are used to
control the orientation of the bead
and magnetic field gradients to apply forces on the bead.
By such techniques the molecule is stretched and twisted and the extension
of the molecule is monitored by the location of the
bead. A typical experimental run measures the extension
of the molecule as a result of the applied twist and force. These
experiments
are important to understanding the role of twist elasticity in
biological processes like DNA replication.

In laboratory and biological conditions the thermal fluctuations
of the DNA molecule are important and it is necessary to take
these into account to understand the experiments
of Strick et al. To determine the response of a DNA molecule to
twist and stretch one has to calculate its partition function:
sum over all possible configurations of the molecule with
Boltzmann weight $\exp[-E/kT]$. As Strick et al remark 
that is a tall order, not likely to be filled anytime soon.
Computer simulations of the system
are possible and have been
done. These simulations,
which incorporate realistic features like self-avoidance,
agree well with laboratory
experiments and
can be viewed in two ways. From the point of view of a laboratory
experimenter they are
like theoretical models. From a theorist's standpoint simulations
can be viewed as
controlled experiments.
The main theoretical difficulty
is the incorporation of self-avoidance, which is present in
a real DNA molecule: the conformation of the molecule must never intersect
itself. Such a constraint is virtually impossible to handle theoretically,
and one is forced to resort to simpler models.
Nevertheless, researchers have attempted to model such a system by 
incorporating a twist term in the energy functional and trying to 
explain the results of the experiments . 
Although the agreement with the experimental curves is good 
there are some issues related to self-avoidance of the polymer 
which are still poorly understood and one needs to introduce an adjustable
cutoff parameter to fit the experimental curves.
The significance of this cutoff parameter is not entirely clear
and there are open issues yet to be understood.

{\bf Finite Size Effects in Single Molecule Experiments}

In the past, experiments on biopolymers were confined to studying their
bulk properties, which involved probing large numbers of molecules.
The results of these experiments could be analyzed by using
the traditional tools of thermodynamics.
However, when one micromanipulates single polymer molecules the usual
rules of thermodynamics applicable to bulk systems no longer hold.
Single molecule experiments thus provide physicists with a concrete testing
ground for understanding some of the fundamental ideas of
statistical mechanics.
In particular, fluctuations about the mean value of a variable
play an important role due to the finite extent of the molecule.
For instance, it turns out, that an experiment in which the distance
between the ends of
a polymer molecule is fixed (an isometric setup) and the
tension fluctuates yields a different
``equation of state'' from one in which the 
tension between the ends is held fixed
(an isotensional setup) and
the end-to-end distance fluctuates.
This asymmetry can be traced to
large fluctuations about the mean value of the force or the extension,
depending on the experimental setup. These fluctuations
vanish only in the thermodynamic limit of very long polymers.
A similar effect is seen in experiments involving twisting DNA molecules.
If the DNA is short enough, experiments with a fixed torque and a fluctuating
twist give a different Torque-Twist Relation from 
those with a fixed twist and a fluctuating torque.

{\bf Equilibrium and Nonequilibrium Statistical Mechanics: A Connection}

We will now come to an interesting connection between equilibrium and 
nonequilibrium statistical
mechanics which has emerged in the past few years and its relevance
to single molecule experiments.

In recent years Jarzinsky has proved a remarkable equality relating work
done in a series of finite time, {\it nonequilibrium} measurements and
the {\it equilibrium} free energy difference between two given configurations
of a system. What is most remarkable is, unlike linear response and the
fluctuation-dissipation theorem (See Box $3$)
which relate equilibrium statistical mechanics
to nonequilibrium processes close to equilibrium, this relation is valid
arbitrarily far from equilibrium.

Let us consider a finite classical system that depends on an external parameter $\lambda$.
Let the system come to equilibrium with a reservoir at a temperature $T$.
If we now switch the external parameter from an initial value $\lambda =0$ to a final
value $\lambda =1$ {\it infinitely} slowly then the work $W_{\infty}$ done on the system is
given by:
$$W_{\infty}= \Delta F = F_1 - F_0$$
where $F_\lambda$ is the  Free energy of the system at a temperature $T$ at a fixed value
$\lambda.$
Let us now consider switching the parameter $\lambda$ at a finite rate. In such a situation
the system is no longer in quasistatic equilibrium with the reservoir at every stage.
The work done in such a case {\it depends} on the microscopic initial conditions of the
system and the reservoir. One now needs to consider 
an {\it ensemble} of such switching
measurements and the total work done is no longer equal 
to the free energy difference but always exceeds it.

It turns out, however, that a remarkable
equality still holds connecting the work done
in such a process and the difference
between the initial and final values of the free
energy. The equality, known as Jarzynski's equality is given by:
${\overline{exp(-\beta W)}} =  exp(-\beta \Delta F)$.
The overbar on the left pertains to
an ensemble average over all possible finite time
measurement realizations.

Recently Liphardt and collaborators have tested the validity of this remarkable equality via
single molecule experimental measurements. They tested the equality by comparing work
done in reversible and irreversible unfolding of a single RNA molecule.

Small beads were attached at the ends of a single RNA molecule. The bead at one end was
held in an optical trap. By measuring the displacement $\delta$ of the bead in the trap and the stiffness constant $k$ of the trap, the force $F=k\delta$ on the molecule was
measured. The bead at the other end was held by a micropipette and attached to a
piezoelectric device which enabled the position of the bead to be controlled by varying
the voltage. The position $z$ of the second bead relative to a given reference point was
used to measure the extent of unfolding of the molecule. The work $W$ done was measured
by integrating the product of the force and the displacement and plotted as a function of
the extension $z$.

The experiment was repeated at varying rates of pull. There were essentially
two classes of measurements: quasistatic, near equilibrium measurements where
the work done was reversible and given by the free energy difference between
the two end values of the control parameter. The other class consisted of
measurements carried out at faster rates of pull
(truly nonequilibrium situation) where the molecule does not have time to
equilibriate during pulling. The experimenters took an average over all
possible realizations of the nonequilibrium measurements and
evaluated the quantity ${\overline{exp(-\beta W)}}$ and  compared it
against its equilibrium
counterpart $exp(-\beta \Delta F)$ and found good agreement within
experimental error
corroborating Jarzynski's equality.

{\bf Concluding Remarks:}

In this article we have tried to give a perspective on the growing and exciting 
area of biopolymer elasticity where there has been a merging of ideas
between biologists, chemists and theoretical physicists. On one hand,
elasticity experiments on biopolymers triggered by a need to understand 
its role in complex biological processes such as transcription and gene
regulation have inspired physicists to come up with models that capture
the essence of the problem and at the same time give quantitative explanations 
for force-extension and twist-extension measurements. On the other hand, 
predictions made by theoretical physicists have suggested new experiments.
Computer simulations have played a two fold role: they have provided researchers
with results of ``controlled experiments'' which are ideal for theoretical
model building. At the same time, the results of the simulations when tested
against real experiments enable experimenters to identify the key factors
which control the experimental results.   
Apart from biological relevance, biopolymer elasticity 
experiments have provided a testing ground for theoretical physicists 
for deeper issues related to connection between equilibrium and nonequilibrium
statistical mechanics. 

{\bf BOX I: Mean Square End-To-End Separation Of A Randomly Lying Cord}

Typical chromosomal DNA molecules are long (contour length $L\approx 50 mm$) 
and when viewed under a microscope look 
like random coils and the statistics of their
ensemble of equilibrium configurations is like that of a random walk. The
mean square distance between the ends of a molecule grows in proportion to 
its contour length $L$:
$$<R^2>= 2 L_p L.$$
The constant $L_p$ is the persistence length which is the length over 
which the DNA molecule can be considered to be approximately straight. 

In general, for a semiflexible polymer, the mean square separation between
its ends predicted by the most popular theoretical model of biopolymer
elasticity, the Worm Like Chain model is:
$$<R^2>= 2 L_p^2[L/L_p + e^{-L/L_p} -1]$$
which in the limit of $L/L_p>>1$ as for a DNA, goes as 
$$<R^2>= 2 L_p L.$$
In the other limit, that is, for a stiff biopolymer like an Actin 
filament
the mean square separation between
ends grows ballistically with the contour length $$<R^2> \sim L^2$$.  

There is an experiment that one can do using a fairly low-tech 
setup to check the statistics of end-to-end separation of polymers. 
This `scaled up' experiment tells us that the statistics of separation 
between the ends of a cord is the same as one gets for a semiflexible 
polymer. The original experiment due to Lemons and Lipscombe
(See Suggested Reading List) consisted
of taking some short ($\approx 0.1 m $) and some long ($\approx 0.8 m$)
segments of a bungee cord of length ($= 0.8 m $) and persistence length
($\approx 0.25 m$). To mimic the effect of randomization, the cord was
dropped onto the floor from a height of $2 m$. Positions were marked
off on the cord corresponding to different segment lengths $L$. 
After dropping the cord, the distance $R$ of separation between the $L=0$
end and one of these marks was measured and for each such segment length the 
procedure was repeated about $40$ times. $<R^2>$ was then plotted against
$L$ and the results agreed well with the theoretical expression 
$$<R^2>= 2 L_p^2[L/L_p + e^{-L/L_p} -1]. $$
The persistence length or the rigidity parameter $L_p$ was determined
from the fit of the data against the theoretically predicted form. 

{\bf BOX 2: Link=Twist+Writhe}

DNA can form a closed loop. Since it is a double stranded helix, the linking
number $Lk$ or the number of times one strand wraps around the other is fixed.
In 1971 Fuller noticed an interesting connection: 
the linking number $Lk$ of a closed ribbon can be 
decomposed into the ``Writhe'' $Wr$ of its backbone plus a 
locally determined ``Twist''$Tw$ :
$$Lk = Tw +Wr.$$
Link $Lk$ counts the number of times that one strand winds around the 
other.
It is an integer and clearly a topological 
quantity and does not change unless 
you cut and 
paste the strands. Writhe $Wr$ is a nonlocal 
quantity which measures the wandering of the tangent vector to the polymer
backbone and it
depends only on the central curve or the backbone of the ribbon
representing the polymer.
Twist $Tw$ is a local quantity that measures the integrated
angular velocity of a strand about the backbone of the polymer.
Both $Tw$ and $Wr$ can change so as to keep $Lk$ unchanged. Fuller noticed 
this 
relation in the context of a question related to supercoiled double stranded
DNA rings raised by J. Vinograd, a molecular biologist. This is an example 
where molecular biology has stimulated 
a mathematical line of enquiry. 

{\bf BOX 3: Fluctuation-Dissipation Theorem}

When you turn on a radio you can sometimes hear a noise which can be 
traced to the irregular motion of electrons. Nyquist
was the first to recognize the connection between such thermal noise
and the impedance of a resistor across which an irregular voltage difference
is induced due to thermal motion of the electrons. One sees a similar 
connection in the context of Brownian motion ( See S. Ramaswamy's article
mentioned in the Suggested Reading List). The generalized theorem relating
the fluctuations of a physical variable of a system in equilibrium to dissipative processes in the system subjected to an external force driving it slightly
away from equilibrium, is called, the Fluctuation-Dissipation Theorem. 
The Fluctuation-Dissipation Theorem works in the linear response regime in which
the response of a system to an external stimulus is directly proportional 
to the strength of the stimulus. 

{\bf Suggested Reading:}

\begin{enumerate}
\item F. B. Fuller, {\it Biophysical Journal} {\bf 68}, 815 (1971).
\item M. Doi and S. F. Edwards, The Theory of Polymer Dynamics, 
(Clarendon, Oxford, 1992).
\item J.Marko and E.D.Siggia, Macromolecules {\bf 28}, 8759(1995).
\item T. R. Strick et al, {\it Science}
{\bf 271}, 1835 (1996).
\item C. Bouchiat et al, {\it Biophysical Journal} {\bf 76}, 409 (1999).
\item  C. Bustamante et al,
Current Opinion in Structural Biology {\bf 10}, 279 (2000).
\item S. Ramaswamy, {\it Resonance} {\bf 5}, 16 (2000).
\item D. S. Lemons and T. C. Lipscombe, {\it Am. J. Phys} {\bf 70}, 570 
(2002).
\item J. Liphardt et al,
{\it Science} {\bf 296}, 1832 (2002).
\item  D. Keller, D. Swigon and C. Bustamante,
{\it Biophysical Journal} {\bf 84}, 733 (2003).

\vbox{
\vspace{0.5cm}
\begin{figure}
\epsfxsize=10.0cm
\epsfysize=6.0cm
\epsffile{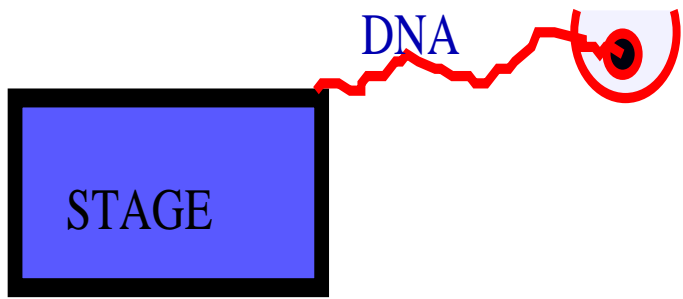}
\caption{Schematic Experimental Setup for a Typical Force Extension 
Measurement.}
\label{expt}
\end{figure}}

\vbox{
\epsfxsize=8.0cm
\epsfysize=6.0cm
\epsffile{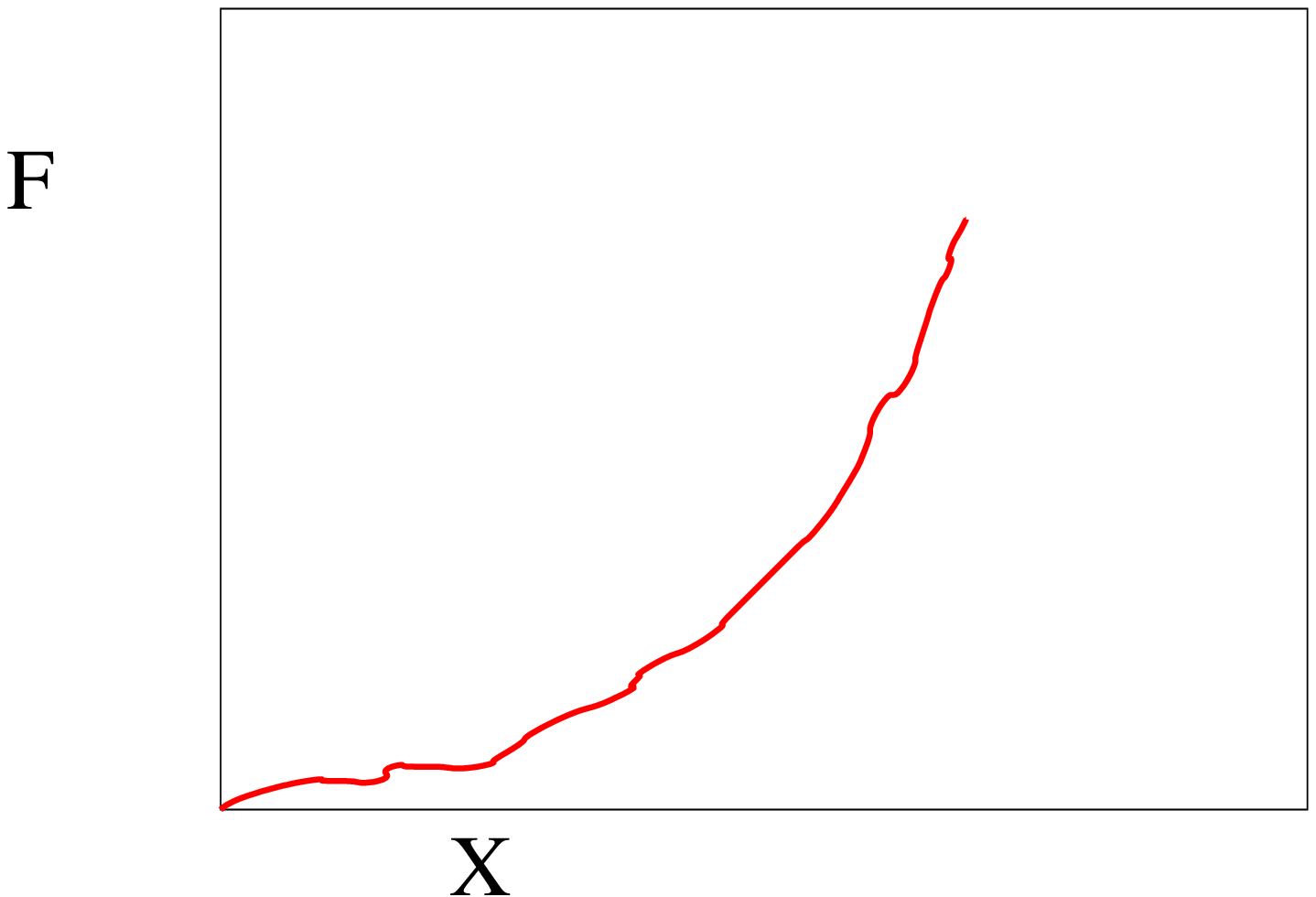}
\begin{figure}
\caption{ Schematic Sketch Of A Typical Force(F)-Extension(X) Curve}
\label{fext}
\end{figure}}

\end{enumerate}

\end{document}